\documentclass[12pt]{article}

\usepackage{amssymb}
\usepackage{amsmath}
\usepackage{amsfonts}

\newcommand{\R}{\mathbb{R}}

\newcommand{\be}{\begin{equation}}
\newcommand{\bea}{\begin{eqnarray}}
\newcommand{\eea}{\end{eqnarray}}

\newcommand{\kt}{\rangle}
\newcommand{\br}{\langle}

\newcommand{\ed}{\end{document}}

\begin{document}

\title{Time-Dependent Hilbert Spaces, Geometric Phases, and
General Covariance in Quantum Mechanics}
\author{\\
Ali Mostafazadeh\thanks{E-mail address: amostafazadeh@ku.edu.tr}\\
\\ Department of Mathematics, Ko\c{c} University,\\
Rumelifeneri Yolu, 34450 Sariyer,\\
Istanbul, Turkey}
\date{ }
\maketitle
\begin{abstract}
We investigate consequences of allowing the Hilbert space of a
quantum system to have a time-dependent metric. For a given
possibly nonstationary quantum system, we show that the
requirement of having a unitary Schr\"odinger time-evolution
identifies the metric with a positive-definite (Ermakov-Lewis)
dynamical invariant of the system. Therefore the geometric phases
are determined by the metric. We construct a unitary map relating
a given time-independent Hilbert space to the time-dependent
Hilbert space defined by a positive-definite dynamical invariant.
This map defines a transformation that changes the metric of the
Hilbert space but leaves the Hamiltonian of the system invariant.
We propose to identify this phenomenon with a quantum mechanical
analogue of the principle of general covariance of General
Relativity. We comment on the implications of this principle for
geometrically equivalent quantum systems and investigate the
underlying symmetry group.
\end{abstract}


\section{Introduction}
Much of our understanding of the universe relies on General
Relativity (GR) and Quantum Mechanics (QM). Besides their
overwhelming success in describing a variety of physical phenomena
and their incredible predictive power, these theories do not
actually have much in common. The lack of a consistent unification
of GR and QM may be linked to their drastic differences in
structure. This point of view underlines the significance of the
study of the basic similarities between these theories and the
search for their alternative formulations that make these
similarities more transparent.

Unlike GR that is a fundamentally geometric theory, QM is
algebraic in nature. Given the appeal of a geometric description,
many researchers have attempted to formulate QM in a geometric
language \cite{geo-qm}. This has also been considered as a natural
prerequisite for embedding QM into a nonlinear theory
\cite{nonlinear-qm} that would share similar geometric features
with GR and hopefully allow for their unification \cite{minic}.
Another development that has led to great interest in exploring
geometric features of QM is the discovery of the geometric phase
and its implications \cite{GP}.

QM is based on a linear vector space ${\cal H}$ equipped with a
(positive-definite) inner product such that the corresponding
metric space is separable and complete. This makes ${\cal H}$ a
sparable Hilbert space whose structure is unique. The situation is
completely opposite in GR. GR is based on a 4-dimensional manifold
$M$ whose structure is neither unique nor fixed by the postulates
of the theory. In GR a concrete physical system corresponds to a
solution of Einstein's field equations. These identify a
pseudo-Riemannian metric on $M$ which is however subject to active
transformations associated with the diffeomorphisms of $M$
\cite{wald}. This is widely referred to as the principle of
general covariance. It is probably the most important feature of
GR and at the same time the very origin of almost all the
difficulties associated with its quantization \cite{carlip}.

The purpose of this article is to reveal a previously unnoticed
similarity of QM to GR, namely that one can consider formulating
the description of a quantum system using a Hilbert space which
does not have a fixed metric and that in doing so one is led to a
symmetry of the system that is associated with certain
transformations of the metric. This symmetry shares the basic
properties of the diffeomorphism-invariance of GR and may be
viewed as a quantum mechanical analogue of the principle of
general covariance.\footnote{A quantum mechanical principle of
general covariance has been advocated in \cite{anandan} that is
different from the one discussed in this paper.}

The article is organized as follows. In Section~2, we review some
basic facts about unitary-equivalent Hilbert spaces. In Section~3,
we outline a unitary quantum description of a given system using a
time-dependent Hilbert space and show that the metric coincides
with a positive-definite dynamical invariant. Furthermore, we
construct a unitary operator that maps the initial
time-independent Hilbert space to the time-dependent Hilbert space
defined by a dynamical invariant and show that the Hamiltonian of
the system is left invariant under this transformation of the
metric. This signifies a symmetry of any quantum system that we
identify as an analogue of the diffeomorphism-invariance of GR. In
Section~4, we show that indeed the above symmetry is associated
not with a particular quantum system but with classes of
geometrically equivalent quantum systems. Here we also comment on
the implications of our findings for a particular method of
generalizing QM. Finally, in Section~5, we discuss the underlying
symmetry group of the quantum general covariance.

In order to concentrate on the conceptual (rather than technical)
issues, we will consider quantum systems with a finite-dimensional
Hilbert space ${\cal H}$. The infinite-dimensional case may be
treated similarly.

\section{Changing the Metric of the Hilbert Space}

In the standard canonical formulation of QM, a quantum system is
uniquely determined by a Hilbert space ${\cal H}$ and a linear
operator $H:{\cal H}\to{\cal H}$ called the Hamiltonian. The
(pure) states of the system are the rays in the Hilbert space
(i.e., the elements of the projective Hilbert space) \cite{nova}.
The physical observables are identified with the self-adjoint
linear operators $O:{\cal H}\to{\cal H}$ and the dynamics is
governed by the Schr\"odinger equation
    \be
    i\hbar\frac{d}{dt}\psi(t)=H\psi(t),
    \label{sch-eq}
    \end{equation}
where $t\in\R$ denotes the time, $\psi(t)\in{\cal H}$ is the
evolving state vector, and the Hamiltonian $H$ may be
time-dependent.

Besides defining the dynamics of the system via the Schr\"odinger
equation (\ref{sch-eq}), the Hamiltonian $H$ also determines the
energy levels and definite-energy (eigen)states of the system. In
order to distinguish the role of the Hamiltonian as the generator
of dynamics and as an observable containing the information about
the energy of the system, it is useful to introduce an energy
observable ${\cal E}:{\cal H}\to{\cal H}$ that coincides with the
Hamiltonian, unless a time-dependent transformation is performed
on the Hilbert space, \cite{energy-op}.

The above description of QM relies on the choice of a fixed metric
(inner product) on ${\cal H}$. The observables (including ${\cal
E}$) are required to be self-adjoint, so that their measured
(eigen)values be real and their eigenvectors that represent the
states of the system after a measurement be orthogonal. The
Hamiltonian $H$ is also assumed to be self-adjoint so that the
time-evolution be unitary. This follows from the condition that
the quantum system admits a probabilistic interpretation in which
the total probability (namely unity) is preserved.

Clearly, the metric of the Hilbert space is not an observable
quantity. It does not appear in the Schr\"odinger
equation~(\ref{sch-eq}) either. Therefore, as far as the physical
content of a quantum system is concerned, the choice of a metric
on ${\cal H}$ is not unique.

It is usually argued that demanding the self-adjointness of a
sufficiently large number of independent observables (a complete
or irreducible set of such) uniquely determines the metric on the
Hilbert space up to an irrelevant multiplicative constant,
\cite{quasi}. This point of view relies on a (quantization) scheme
in which a complete set of observables may be identified with
certain linear operators acting in a vector space of state
vectors. The above requirement of self-adjointness then leads to
an essentially fixed choice for the metric on this space. This
however does not preclude the possibility of a simultaneous
transformation of the metric, the Hamiltonian, and the observables
in such a way that the transformed quantities yield an equivalent
description of the same system. We will identify such a
transformation with a {\em symmetry} of the system, if it leaves
the Schr\"odinger operator $i\hbar\partial_t-H$ invariant. For a
time-independent transformation this condition reduces to the
invariance of the Hamiltonian $H$, \cite{nova}

The above definition of symmetry is slightly more general than the
one that is usually adopted in quantum mechanics. The latter
corresponds to imposing the additional condition that the symmetry
transformations belong to the group $U({\cal H})$ of the unitary
transformations of the Hilbert space, i.e., in conventional
approach to QM, a symmetry transformation is a metric-preserving
linear operator that acts in the Hilbert space and commutes with
the Schr\"odinger operator $i\hbar\partial_t-H$, \cite{nova}; in
particular, a time-independent symmetry transformation is a
metric-preserving linear operator acting in the Hilbert space and
commuting with the Hamiltonian.

The apparent freedom in the choice of the metric on ${\cal H}$ is,
however, overshadowed by the fact that any two metrics on ${\cal
H}$ lead to the same Hilbert space structure. In particular, one
may relate them by a unitary operator.\footnote{A linear operator
${\cal U}:{\cal H}_1\to{\cal H}_2$ relating two Hilbert spaces
${\cal H}_1$ and ${\cal H}_2$ is said to be a unitary operator or
an isometry if for all $\psi,\phi\in{\cal H}_1$, $\br{\cal
U}\psi,{\cal U}\phi\kt_2=\br\psi,\phi\kt_1$, where
$\br\cdot,\cdot\kt_1$ and $\br\cdot,\cdot\kt_2$ respectively stand
for the inner product of ${\cal H}_1$ and ${\cal H}_2$.}

Now, consider a quantum system $S$ whose kinematical and dynamical
aspects are respectively described by a Hilbert space ${\cal H}_1$
and a Hamiltonian operator $H_1:{\cal H}_1\to{\cal H}_1$.
Furthermore, let ${\cal H}_2$ be another Hilbert space, such that
${\cal H}_1$ and ${\cal H}_2$ are unitarily equivalent, i.e.,
there is a (possibly time-dependent) unitary operator ${\cal
U}:{\cal H}_1\to{\cal H}_2$. Then the system $S$ may also be
described by the Hilbert space ${\cal H}_2$ and the Hamiltonian
operator
    \be
    H_2:={\cal U}H_1{\cal U}^{-1}-i\hbar\,{\cal U}\;\frac{d}{dt}\;
    {\cal U}^{-1}.
    \label{H=H}
    \end{equation}
The observables $O_2$ associated with the Hilbert space ${\cal
H}_2$ are related to the observables $O_1$ associated with ${\cal
H}_1$ according to
    \be
    O_2={\cal U}\;O_1{\cal U}^{-1}.
    \label{O=O}
    \end{equation}

Relations~(\ref{H=H}) and (\ref{O=O}) imply that ${\cal U}$ maps
the solutions $\psi_1(t)$ of the Schr\"odinger equation defined by
the Hamiltonian $H_1$ to the solutions $\psi_2(t)$ of the
Schr\"odinger equation defined by $H_2$, i.e., $\psi_2(t)={\cal
U}\psi_1(t)$, and that it leaves the transition amplitudes between
the energy states and the expectation values of the observables
invariant. Therefore, the descriptions of $S$ in terms of $({\cal
H}_1,H_1)$ and $({\cal H}_2,H_2)$ are equivalent. The fact that
the energy operator ${\cal E}_2={\cal U}{\cal E}_1{\cal U}^{-1}$
differs from the transformed Hamiltonian $H_2$ is actually
necessary for the validity of this equivalence.\footnote{If ${\cal
H}_1={\cal H}_2$, we can use any element of the group of all
unitary operators acting in ${\cal H}_1$ to perform the above
transformation of the Hamiltonian and the observables. By
definition, these transformations leave the inner product (and
metric) of the Hilbert space ${\cal H}_1$ invariant. They
correspond to (time-dependent) quantum canonical transformations
\cite{nova}.}

Next, suppose ${\cal H}_1$ and ${\cal H}_2$ are the Hilbert spaces
obtained by endowing a vector space ${\cal V}$ with the inner
products $\br\cdot,\cdot\kt_1$ and $\br\cdot,\cdot\kt_2$
respectively. Then there is a positive-definite
operator\footnote{A positive-definite operator is a positive
invertible operator. Alternatively, it is a self-adjoint operator
with a positive spectrum.} $\eta:{\cal H}_1\to{\cal H}_1$, called
a {\em metric}, such that for all $\psi,\phi\in{\cal V}$,
\cite{kato},
    \be
    \br\psi,\phi\kt_2=\br\psi,\eta\phi\kt_1.
    \label{e1}
    \end{equation}
Because $\eta$ is a positive-definite operator, it has a unique
positive-definite square root $\rho:{\cal H}_1\to{\cal H}_1$. If
we view $\rho^{-1}$ as an operator mapping ${\cal H}_1$ to ${\cal
H}_2$, we can easily check that it is unitary: for all
$\psi,\phi\in {\cal H}_1$,
    \[ \br\rho^{-1}\psi,\rho^{-1}\phi\kt_2=
        \br\rho^{-1}\psi,\eta\rho^{-1}\phi\kt_1=
        \br\psi,\rho^{-1\dagger}\eta\rho^{-1}\phi\kt_1=
        \br\psi,\phi\kt_1.\]
Here and in what follows $\dagger$ denotes the adjoint of an
operator viewed as acting in the Hilbert space ${\cal H}_1$, i.e.,
for a linear operator $L$ acting in ${\cal V}$, $L^\dagger$ is
defined by $\br\psi_1,L\psi_2\kt_1=\br
L^\dagger\psi_1,\psi_2\kt_1$ for all $\psi_1,\phi_1\in{\cal V}$.

This completes the demonstration of the unitary-equivalence of any
two Hilbert spaces with the same vector space structure
(dimension) \cite{simon-reed,nik}. Its direct consequence is that
one does not gain much by using different inner products. This is
certainly true, if one restricts oneself to time-independent
metric operators $\eta$ and the corresponding inner products.

\section{Time-Dependent Hilbert Spaces and Dynamical Invariants}

The need for formulating quantum mechanics on a Hilbert space with
a time-dependent inner product arises in, for example, trying to
develop a nonrelativistic quantum mechanics of a particle confined
to move on an oscillating membrane or a particle subject to a
time-dependent inhomogeneous gravitation field such as a
gravitational wave.\footnote{The same is true of studying the
motion of a nonrelativistic particle confined to a box with moving
walls. The standard approach to this problem is to make a unitary
transformation that fixes the the Hilbert space but makes the
potential time-dependent \cite{seba}. For further details and
references see \cite{nova}.} Nevertheless, to the author's best
knowledge, a comprehensive treatment of quantum mechanics with a
time-dependent Hilbert space has not yet appeared. The subject has
been considered within the context of canonical quantum gravity
where the idea is to amend the Schr\"odinger equation with an
extra term to compensate for the contribution of the metric
operator that renders the time-evolution non-unitary even for a
self-adjoint Hamiltonian \cite{isham}.\footnote{For a similar
approach see \cite{box}.} This is a rather drastic departure from
the ordinary QM. We shall pursue an alternative approach that
allows for a unitary time-evolution without modifying the
Schr\"odinger equation. It has its roots in a recent attempt to
resolve some of the basic problems of quantum cosmology
\cite{cqg}.

Let ${\cal H}_1$ and ${\cal H}_2$ be the Hilbert spaces obtained
by endowing a vector space ${\cal V}$ with a time-independent
inner product $\br\cdot,\cdot\kt_1$ and a time-dependent inner
product $\br\cdot,\cdot\kt_2$, respectively. According to
(\ref{e1}), $\br\cdot,\cdot\kt_2$ may be expressed in terms of
$\br\cdot,\cdot\kt_1$ and a time-dependent metric operator
$\eta=\eta(t)$. Then a linear operator $H:{\cal V}\to{\cal V}$
defines a unitary time-evolution in ${\cal H}_2$ according to the
Schr\"odinger equation~(\ref{sch-eq}), if and only if for any pair
of solutions $\psi_2(t)$ and $\phi_2(t)$ of~(\ref{sch-eq}) we have
    \be
    \frac{d}{dt}\,\br\psi_2(t),\phi_2(t)\kt_2=0.
    \label{e2}
    \end{equation}
Substituting (\ref{e1}) in this equation yields
    \be
    i\hbar\frac{d}{dt}\,\eta=H^\dagger\eta-\eta H.
    \label{e3}
    \end{equation}
In particular if $H$ is self-adjoint with respect to the inner
product $\br\cdot,\cdot\kt_1$, then we obtain
    \be
    i\hbar\frac{d}{dt}\,\eta=[H,\eta].
    \label{dyn-inv}
    \end{equation}
This is the defining (Liouville-von~Neumann) equation for a
dynamical invariant \cite{lewis-riesenfeld,nova} for a
self-adjoint Hamiltonian $H$ acting in ${\cal H}_1$.
Equation~(\ref{e3}) is its generalization for a non-self-adjoint
Hamiltonian.\footnote{Note however that the dynamical invariants
obtained in this was are positive-definite operators.}

Indeed, the most general solution of (\ref{e3}) is given by
\cite{cqg,qc}
    \be
    \eta(t)=U(t,t_0)^{-1\dagger}\eta_0U(t,t_0)^{-1},
    \label{e5}
    \end{equation}
where
    \be
    U(t,t_0):={\cal T}\; e^{-\frac{i}{\hbar}\int_{t_0}^t H(t')dt'}
    \label{U=}
    \end{equation}
is the time-evolution operator for the Hamiltonian $H$, ${\cal T}$
is the time-ordering operator, $t_0\in\R$ is an initial time, and
$\eta_0:{\cal H}_1\to{\cal H}_1$ is a time-independent
positive-definite operator. If we suppose that, as an operator
acting in ${\cal H}_1$, $H$ is self-adjoint, i.e., $H^\dagger=H$,
then (\ref{U=}) reduces to the following familiar relation for the
dynamical invariants \cite{nova}
    \be
    \eta(t)=U(t,t_0)\eta_0U(t,t_0)^\dagger.
    \label{e6}
    \end{equation}
In this case the restriction of the positive-definiteness of
$\eta_0$ is actually not significant. Given any dynamical
invariant $I(t)$ for a self-adjoint Hamiltonian $H:{\cal
H}_1\to{\cal H}_1$, one can easily construct a positive definite
invariant, e.g., if we use $1$ to denote the identity operator for
${\cal H}_1$, then $I^2+1$ is a positive-definite invariant.

The requirement that $H$ be self-adjoint with respect to the inner
product $\br\cdot,\cdot\kt_1$ of ${\cal H}_1$ raises the following
question. Given that we wish to formulate a quantum mechanics
using the time-dependent Hilbert space ${\cal H}_2$, should not we
require the Hamiltonian $H$ to be self-adjoint with respect to the
inner product $\br\cdot,\cdot\kt_2$? The answer to this question
is actually negative. One can check that adopting any metric
operator of the form (\ref{e5}) ensures the unitarity of the
time-evolution. Therefore, in principle, the Hamiltonian $H$ need
not be self-adjoint with respect to $\br\cdot,\cdot\kt_2$.
However, it turns out that requiring $H$ to be self-adjoint with
respect to $\br\cdot,\cdot\kt_1$ does imply that it is also
self-adjoint with respect to $\br\cdot,\cdot\kt_2$. To see this,
we view $H$ as a self-adjoint Hamiltonian acting in the Hilbert
space ${\cal H}_1$ and try to use the unitary map
$\rho^{-1}=\eta^{-1/2}$ to obtain the transformed Hamiltonian
$H_2$ acting in ${\cal H}_2$. In view of (\ref{e6}),
    \be
    \rho(t)^{-1}=U(t,t_0)\eta_0^{-1/2}U(t,t_0)^\dagger.
    \label{e7}
    \end{equation}
Hence $\rho(t)^{-1}$ is also a positive-definite dynamical
invariant. Substituting $H$ for $H_1$ and $\rho^{-1}$ for ${\cal
U}$ in (\ref{H=H}), we find the rather remarkable result
    \be
    H_2=H.
    \label{e8}
    \end{equation}
In other words, the transformation induced by the unitary operator
$\rho^{-1}$ changes the metric of the Hilbert space into a
time-dependent metric, but it leaves the Hamiltonian of the system
invariant. Because by construction $\rho^{-1}$, viewed as mapping
${\cal H}_1$ onto ${\cal H}_2$, is a unitary operator and $H_1=H$
is assumed to be self-adjoint with respect to
$\br\cdot,\cdot\kt_1$, the Hamiltonian $H=H_2$ is also
self-adjoint with respect to $\br\cdot,\cdot\kt_2$.

Unlike the Hamiltonian $H$, the observables and in particular the
energy operator ${\cal E}$ do change under the unitary
transformation induced by $\rho^{-1}$. However, their expectation
values which are of physical significance remain invariant. As a
result $\rho^{-1}$ is a genuine symmetry transformation for the
physical system.  It differs from the ordinary symmetry
transformations --- that are linked with the degeneracy structure
of the Hamiltonian and realized in terms of the unitary
transformations acting within a fixed Hilbert space \cite{nova}
--- in that it changes the metric of the Hilbert space.

In a sense, the ordinary symmetries are the analogues of the
passive coordinate transformations of GR and the metric-changing
symmetries such as the ones induced by the invariants $\rho^{-1}$
are the analogues of the active diffeomorphisms of GR. In view of
this analogy it is tempting to refer to the presence of the
above-described metric-changing symmetries of a quantum system as
a quantum mechanical principle of general covariance. The
principle of general covariance of GR stems from the symmetry of
the Einstein's field equation (alternatively of the
Hilbert-Einstein action) under active diffeomorphisms of the
spacetime manifold. The quantum mechanical general covariance may
also be viewed as a consequence of the invariance of the
Schr\"odinger equation under the metric-changing symmetry
transformations of the Hilbert space.

In summary, we identified a simple but generic symmetry of the
quantum mechanical description of an arbitrary physical system.
The corresponding symmetry transformations are defined by the
positive-definite dynamical invariants. They change the metric of
the Hilbert space but leave all the physical quantities associated
with the system invariant. We propose to refer to this invariance
or symmetry principle as the quantum mechanical general
covariance.

\section{Geometrically Equivalent Quantum Systems and
Geometric Phases}

Consider a pair of quantum systems $S_1$ and $S_2$ that are
respectively described by the Hilbert spaces ${\cal H}_1$ and
${\cal H}_2$ and the Hamiltonians $H_1$ and $H_2$. Then, by
definition, $S_1$ and $S_2$ are said to be {\em geometrically
equivalent} if ${\cal H}_1={\cal H}_2$ and $S_1$ and $S_2$ share
identical geometric phases for a complete set of initial states,
alternatively they share a common nontrivial dynamical invariant
\cite{jpa-2001b}. As we argued in Section~3, one can always
construct a nontrivial positive-definite dynamical invariant out
of a given nontrivial invariant.\footnote{Here by a nontrivial
invariant, we mean a solution of (\ref{dyn-inv}) that is not a
multiple of the identity operator.}

Let $\eta$ be a common positive-definite dynamical invariant of a
pair of geometrically equivalent quantum systems $S_1$ and $S_2$.
Then the unitary transformation defined by
$\rho^{-1}:=\eta^{-1/2}$, that changes the fixed metric of the
Hilbert space into the time-dependent metric $\eta$, leaves both
the Hamiltonians $H_1$ and $H_2$ invariant. Therefore, the
metric-changing symmetry-transformations actually signify the
symmetries of the geometrically equivalent Hamiltonians. The
latter differ by the ordinary symmetries of the invariant
\cite{jpa-2001b}, i.e., $[H_1-H_2,\eta]=0$.

Next, we recall that the basic property of a dynamical invariant
is that its eigenvalue problem is essentially equivalent to the
time-dependent Schr\"odinger equation~(\ref{sch-eq}), in the sense
that it has a complete set of eigenvectors that solve the
Schr\"odinger equation~(\ref{sch-eq}). Therefore, any solution of
(\ref{sch-eq}) may be expressed as a linear combination of a set
of eigenvectors of the invariant \cite{lewis-riesenfeld,nova}.
Specifically, suppose (for similicity) that $\eta$ is a
positive-definite dynamical invariant with a nondegenerate
spectrum. Then it can be shown that the eigenvalues $\lambda_n$ of
$\eta$ are constant and that the evolution operator associated
with the Hamiltonian $H$ may be expressed in the form
    \be
    U(t,t_0)=\sum_n e^{i\alpha_n(t,t_0)}|\lambda_n;t\kt\br
    \lambda_n;t_0|,
    \label{evo-op}
    \end{equation}
where $\{|\lambda_n;t\kt\}$ is any complete set of orthonormal
eigenvectors of $\eta$ and
    \bea
    \alpha_n(t,t_0)&=&\delta_n(t,t_0)+\gamma_n(t,t_0),
    \label{tot-p}\\
    \delta_n(t,t_0)&:=&-\frac{1}{\hbar}\int_{t_0}^t
    \br\lambda_n;t'|H(t')|\lambda_n;t'\kt~dt',
    \label{dyn-p}\\
    \gamma_n(t,t_0)&=&i\int_{t_0}^t
    \br\lambda_n;t'|\frac{d}{dt'}|\lambda_n;t'\kt~dt'.
    \label{geo-p}
    \eea
The phase angles $\delta_n$ and $\gamma_n$ are known as the
dynamical and geometric parts of the total phase angle $\alpha_n$,
\cite{nova}. If for some $T\in\R$,
$|\lambda_n;T+t_0\kt=|\lambda_n;t_0\kt$, then the initial state
corresponding to the state vector $|\lambda_n;t_0\kt$ will be a
cyclic state with $\delta_n(t_0+T,t_0)$ and $\gamma_n(t_0+T,t_0)$
being the corresponding dynamical and (cyclic) geometric phase
angles.

Identifying the positive-definite dynamical invariant $\eta$ with
a metric on the Hilbert space reveals the curious fact that the
{\em geometric phases are determined by the metric}. The latter
also determines the evolving state in the projective Hilbert
space. But it does not provide the information about the dynamical
phase angles, and therefore falls short of fully determining the
dynamics (specifically the evolution operator (\ref{evo-op})) of
the system. This is the key obstruction to formulating QM in terms
of a dynamical metric $\eta(t)$ on the Hilbert space and the
corresponding linear `field equation' (\ref{dyn-inv}). Such a
formulation requires, in addition, a mechanism to specify the
dynamical phase angles $\delta_n$.

The observations described in the preceding paragraph may be
viewed as grounds for a generalization of QM into a theory in
which the dynamics of a physical system is determined by a
time-dependent metric on the Hilbert space together with a
prescription for specifying the dynamical angles $\delta_n$ and a
`field equation' for the metric that would be more general than
(\ref{dyn-inv}). A potential candidate for the latter would be a
master equation of the Lindblad type \cite{lindblad} that is used
in modelling dissipation in QM. This would generalize
(\ref{dyn-inv}) and (\ref{e3}) to
    \be
    i\hbar\left[\frac{d}{dt}\eta+D(\eta)\right]=
    H^\dagger\eta-\eta H,
    \label{lin}
    \end{equation}
where $D$ has the general form
    \[D(\eta)=\frac{1}{2}\sum_{j=1}^r\left([A_j^\dagger,A_j\eta]+
    [\eta A_j^\dagger,A_j]\right),\]
and $A_j$ are the Lindblad operators, \cite{b-g}.

If one adopts the master equation (\ref{lin}) to determine the
metric $\eta$ --- instead of (\ref{dyn-inv}) --- and uses the same
prescription to compute the dynamical phases appearing in
(\ref{evo-op}) as in the ordinary QM, namely using
Eq.~(\ref{dyn-p}), then one arrives at a generalization of the
ordinary QM. The dynamics of the resulting theory is determined
through the action of the time-evolution operator $U(t,t_0)$
according to $|\psi(t)\kt=U(t,t_0)|\psi(t_0)\kt$, where $U(t,t_0)$
is given by (\ref{evo-op}) in which $|\lambda_n;t\kt$ form a
complete set of orthonormal eigenvectors of $\eta$ and the phase
angles $\alpha_n(t,t_0)$ are given by (\ref{tot-p}) --
(\ref{geo-p}). For a Hermitian Hamiltonian (in the original fixed
metric on the Hilbert space ${\cal H}$) $U(t,t_0)$ would again be
unitary. Therefore, it can be linked to a Schr\"odinger equation
with a Hermitian Hamiltonian:
    \[H'(t):=i\hbar\left[\frac{d}{dt}U(t,t_0)\right]
    U(t,t_0)^{-1}.\]
However, as operators acting in the Hilbert space endowed with the
metric $\eta$, $U(t,t_0)$ will not generally be unitary and
$H'(t)$ will not be Hermitian. It is also not difficult to see
that because of the dissipative term $D$ in (\ref{lin}) the
quantum systems defined by $H'(t)$ in the fixed Hilbert space is
not unitarily equivalent to the one defined by $H'(t)$ in the
time-dependent system. The evolution defined by $U(t,t_0)$, as
constructed above, will be generally nonunitary. This calls for a
more detailed investigation of the implications and potential
applications of this type of generalizations of QM that is based
on a time-dependent metric.

\section{Underlying Group Structure}

Consider a quantum system $S$ with a fixed Hilbert space ${\cal
H}$ and a Hamiltonian $H$. Suppose that $\eta_1$ and $\eta_2$ are
a pair of positive-definite dynamical invariants of $H$, and
${\cal H}_i$ is the Hilbert space that has the same vector space
structure as ${\cal H}$ but endowed with the inner product
$\br\cdot,\eta_i\cdot\kt$, for $i\in\{1,2\}$. Then $\eta_i^{-1/2}$
is the unitary operator mapping ${\cal H}$ onto ${\cal H}_i$ and
consequently ${\cal U}:=\eta_2^{-1/2}\eta_1^{1/2}$ is the unitary
operator mapping ${\cal H}_1$ onto ${\cal H}_2$. Clearly, $H$ is
invariant under all these transformations. However, the operator
${\cal U}$ viewed as an operator acting in ${\cal H}$ is not a
positive-definite invariant of the Hamiltonian $H$, for although
it satisfies the defining relation of an invariant it is not
generally Hermitian with respect to the inner product on ${\cal
H}$. This is easily seen if we use (\ref{e6}) to express ${\cal
U}$ in the form
    \[{\cal U}(t)=U(t,t_0)\eta_2(t_0)^{-1/2}\eta_1(t_0)^{1/2}
    U(t,t_0)^\dagger.\]

Because $\eta_1$ and $\eta_2$ are determined by their initial
values $\eta_1(t_0)$ and $\eta_2(t_0)$, we may view ${\cal
U}(t_0)$ as the transformation that maps the metric $\eta_1$ to
$\eta_2$. As any pair of metrics may be related in this way (while
preserving the Hamiltonian), we may identify ${\cal U}(t_0)$ with
an element of the permutation group ${\cal S}_{_{\cal M}}$ of the
set ${\cal M}$ of all positive-definite operators $\eta_0$ acting
in the Hilbert space ${\cal H}$. It is trivial to see that the
converse is also true, i.e., any such permutation may be affected
by an operator of the form ${\cal U}(t_0)$. This shows that the
principle of general covariance associated with the presence of
metric-changing symmetries of a quantum system has the permutation
group ${\cal S}_{_{\cal M}}$ as its underlying symmetry group.
This is the quantum mechanical analogue of the diffeomorphism
group of spacetime in GR.

The main difference is that in GR even after moding out the
diffeomorphism symmetry one still has a continuum infinity of
possible geometries of the spacetime, whereas in QM moding out the
above-mentioned permutation group symmetry one is left with a
unique Hilbert space structure. The latter may be viewed as
`fixing a gauge' that corresponds to a particular metric on the
Hilbert space. The conventional choice for the gauge (metric) is
the usual time-independent one. But choosing a particular gauge
does not destroy the gauge freedom. The main purpose of this
article is to show that one may choose a gauge (a metric) that is
time-dependent. This identifies it with a positive-definite
dynamical invariant and sheds light on a variety of issues related
to geometric phases, geometric descriptions of QM, and ways of
generalizing it.

\section{Conclusion}

The uniqueness of the Hilbert space structure of the Hilbert space
of a quantum system usually leads one to undermine the fact that
the metric of the Hilbert space is not fixed by physical
considerations. Allowing the metric to be dynamical reveals an
interesting connection to the theory of dynamical invariants. It
shows that the geometric phases may be identified as the
properties of a time-dependent metric on the Hilbert space whereas
the dynamical phases are not linked to such a manifestly geometric
structure.

Another simple outcome of our analysis is the rather remarkable
observation that one can indeed view density operators, that form
positive-definite dynamical invariants, also as time-dependent
metric operators.

The freedom in the choice of a metric among the set ${\cal M}$ of
all positive-definite dynamical invariants of a given quantum
system may be viewed as an albeit rather trivial quantum
mechanical analogue of the principle of general covariance of GR.
The role of the diffeomorphism group of GR is played by the
permutation group of ${\cal M}$.

The Schr\"odinger dynamics of a given system may be formulated in
terms of the Liouville-von~Neumann equation (\ref{dyn-inv}) for
the metric, that determines the evolving states as well as the
corresponding geometric phases, and an additional prescription for
computing the dynamical phases. This point of view may be used as
the basis of a class of generalizations of QM in which the
Liouville-von~Neumann equation is replaced by a more general
`field equation' for the metric on the Hilbert space. A nonunitary
class of examples are provided by the Lindblad's master equation.

\subsection*{Acknowledgment}
Eq.~(\ref{e8}) was initially derived by Ahmet Batal as a part of
his independent study project. This work has been supported by the
Turkish Academy of Sciences in the framework of the Young
Researcher Award Program (EA-T$\ddot{\rm U}$BA-GEB$\dot{\rm
I}$P/2001-1-1).

\ed